\documentclass{article}
\usepackage{arxiv}

\usepackage[utf8]{inputenc} 
\usepackage[T1]{fontenc}    
\usepackage{hyperref}       
\usepackage{url}            
\usepackage{booktabs}       
\usepackage{amsfonts}       
\usepackage{nicefrac}       
\usepackage{microtype}      
\usepackage{lipsum}
\usepackage{float}
\usepackage{amsmath, amsfonts}
\usepackage{graphicx}
\usepackage{comment}
\usepackage{xcolor}
\usepackage{subfig}
\usepackage[normalem]{ulem}
\usepackage{array,multirow}
\usepackage{geometry}
\newcolumntype{L}[1]{>{\raggedright\arraybackslash}p{#1}}
\title{Tracking Temporal Evolution of Network Activity for Botnet Detection}

\author{
  Kapil Sinha \\
  Department of Computer Science\\
  California Institute of Technology\\
  Pasadena, CA 15213 \\
  \texttt{ksinha@caltech.edu} \\
   \And
 Arun Viswanathan \\
 Cyber Defense Engineering and Research \\
  Jet Propulsion Laboratory, California Institute of Technology\\
  Pasadena, California\\
  \texttt{aviswant@jpl.nasa.gov} \\
     \And
 Julian Bunn \\
 Center for Data Driven Discovery,\\
 California Institute of Technology\\
 Pasadena, CA 15213 \\
  \texttt{Julian.Bunn@caltech.edu}
}

\begin{document}
\maketitle

\begin{abstract}
Botnets are becoming increasingly prevalent as the primary enabling technology in a variety of malicious campaigns such as email spam, click fraud, distributed denial-of-service (DDoS) attacks, and cryptocurrency mining. 
Botnet technology has continued to evolve rapidly making detection a very challenging problem. 
There is a fundamental need for robust detection methods that are insensitive to characteristics of a specific botnet and are generalizable across different botnet types.
We propose a novel supervised approach to detect malicious botnet hosts by tracking 
a host's network activity over time using a
Long Short-Term Memory (LSTM) based neural network architecture.
We build a prototype to demonstrate the feasibility of our approach, evaluate it on the CTU-13 dataset, and compare our performance against existing detection methods. 
We show that our approach results in a more generalizable, botnet-agnostic detection methodology, is amenable to real-time implementation, and performs well compared to existing approaches, with an overall accuracy score of 96.2\%.
\end{abstract}


\section{Introduction}

Botnets are groups of connected, malware-infected hosts (bots) that can be controlled by a remote attacker. 
They are prominent threats in cybersecurity, often used for purposes ranging from distributed denial-of-service attacks and click-fraud, to email spam and cryptocurrency mining~\cite{cryptocurrency}.
As per recent estimates, botnets control billions of infected hosts worldwide and are responsible for seventy percent of all spam~\cite{background}. 

While botnets have existed for many years, they continue to evolve and become  sophisticated.
Newer botnets often encrypt their packets, 
vary their control protocols, and 
use peer-to-peer topologies rather than centralized ones to improve their robustness~\cite{background}. 
Thus, traditionally used  
signature-based~\cite{signature-based}, heuristic-based~\cite{heuristic-based},
and 
content-based~\cite{content-based} methods for detecting botnets are rendered ineffective and are less generalizable, making detection of previously unseen or newer botnets difficult.

On the other hand, anomaly-based detection methods, as are common in intrusion detection systems, show potential for detecting previously unseen or newer botnets. 
There are two broad anomaly-based approaches seen in literature.
One approach trains on nominal or non-malicious network traffic, and classify any traffic outside of the nominal range as botnet activity~\cite{graph-based}. 
The other approach trains on 
different types of botnet traffic and detects similarly malicious botnet traffic~\cite{ctu13}.

The recent trend in botnet detection has been to apply one of the above anomaly-based approaches over flow-based or graph-based features.
Flow-based methods aggregate network packet data into a sequence of time stamped flows, where a flow is defined as a tuple consisting of source IP address, destination IP address, port, and sometimes protocol. 
They analyze standard packet-level or flow-level statistics such as payload length, mean number of bytes per packet, and flow duration using machine-learning techniques such as supervised classification or clustering~\cite{flow-based}. 
The flow features, however, are often characteristic of specific protocol-based botnets, and are not generalizable to newer botnet varieties. 

Graph-based methods are relatively new in botnet detection~\cite{graph-based, botgrep, social-graph}.
These methods ignore the sequential nature of the data, and focus
on the graphical structure of the communication using 
centrality-based graph measures.
However, they often lend themselves poorly towards an actually useful tool, 
as they require access to all the data at once to build the graph models. 

In this work, we propose a novel supervised anomaly-detection approach by 
modeling graph features over time, 
which results in better detection performance as compared to existing methods,
and lends itself better as a practical, implementable tool.

\paragraph{High-level Overview of Approach}
Our approach examines the communication graph of hosts over time to find malicious hosts. 
We generate graphs of the network communication at regular intervals and extract basic graph-based statistical and centrality features. 
Then, we assemble a time-series of these features for each host (identified by IP address) and 
train a time-series classification model to determine which hosts are most likely to be botnet-infected.

As we employ a graph-based approach, our detection method is agnostic to both packet content and protocol, instead targeting a botnet's communication structure. Hence, attempts to avoid detection via packet encryption and spoofing are ineffective. Moreover, our detection method generalizes to a variety of different botnets as communication structures are often very similar across botnets.

Some recent literature focuses on clustering or filtering nodes in order to narrow down the set of potentially malicious nodes~\cite{graph-based}. Our approach avoids the need for this pre-processing step and is able to identify malicious nodes directly from the set of all nodes in the network. Rather than requiring a large aggregate amount of data, our approach exploits the time series data in the network and thus is more amenable for real-time detection. We discuss details of our methodology in Section~\ref{sec:methodology}.

\paragraph{Evaluation}
We develop a prototype implementation of our algorithm, and evaluate it over the
over the benchmark CTU-13 dataset~\cite{ctu13}.
Our evaluation focuses on scenarios 6, 7, 10, 11, and 12 of the CTU-13 dataset, which  encompass Command \& Control, DDoS, and P2P botnets consisting of 1, 3, or 10 bots.

The existing performance of the  state-of-the-art detector on the CTU-13 dataset achieves a true positive rate (TPR) of $0.809$ and false positive rate (FPR) of $0.030$~\cite{rnn}. 
While our FPR is slightly higher at $0.037$, our approach easily outperforms the state-of-the-art TPR with a $0.946$.
We discuss the details of our evaluation methodology and results in Section~\ref{sec:results}.

\paragraph{Key contributions}
Our paper makes the following key contributions:
(a) we present a generalizable, novel, content-agnostic algorithm for detection of botnet-infected hosts by tracking the temporal evolution of botnet communication structure; and
(b) we perform a comprehensive comparison of our algorithm with previous algorithms evaluated on the CTU-13 dataset to demonstrate that our 
approach outperforms all content-agnostic approaches, and several
content-aware approaches.
Finally, all our source code developed for this project is released as-is in the public domain~\footnote{https://github.com/kapilsinha/botnet-surf}.

\paragraph{Paper Outline} The rest of the paper is structured as follows. 
Section~\ref{sec:algorithm} discusses the botnet detection algorithm.
Section~\ref{sec:results} discusses our detailed evaluation methodology and results on the 
CTU-13 dataset. 
Section~\ref{sec:discussion} presents a comprehensive comparison of our approach to other approaches published on the CTU-13 dataset,
discusses the implementation performance of our algorithm, and 
describes the key limitations of our overall approach.
Section~\ref{sec:conclusion} concludes the paper with a brief discussion of the future work.


\section{Botnet Detection Algorithm} 
\label{sec:methodology}
\label{sec:algorithm}

In this section, we first motivate our algorithm with an example, followed by a discussion of the algorithm workflow and the details of each step.

\subsection{Motivation}

\begin{figure}[!htbp]
\centering
\begin{tabular}{cccc}
\subfloat[t $\in$ [600 s, 720 s)]{\includegraphics[width = 1.4in]{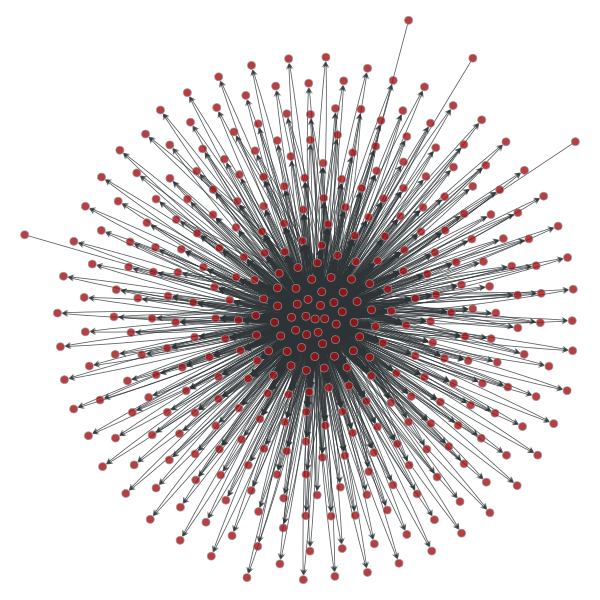}} &
\subfloat[t $\in$ [660 s, 780 s)]{\includegraphics[width = 1.4in]{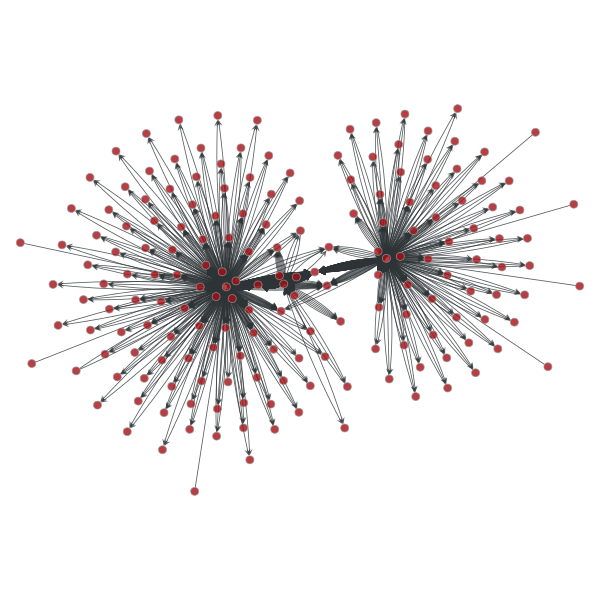}} &
\subfloat[t $\in$ [720 s, 840 s)]{\includegraphics[width = 1.4in]{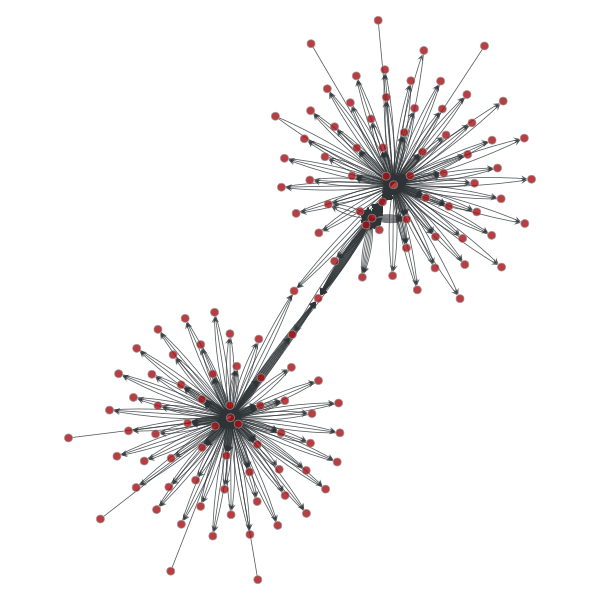}} \\
\subfloat[t $\in$ [780 s, 900 s)]{\includegraphics[width = 1.4in]{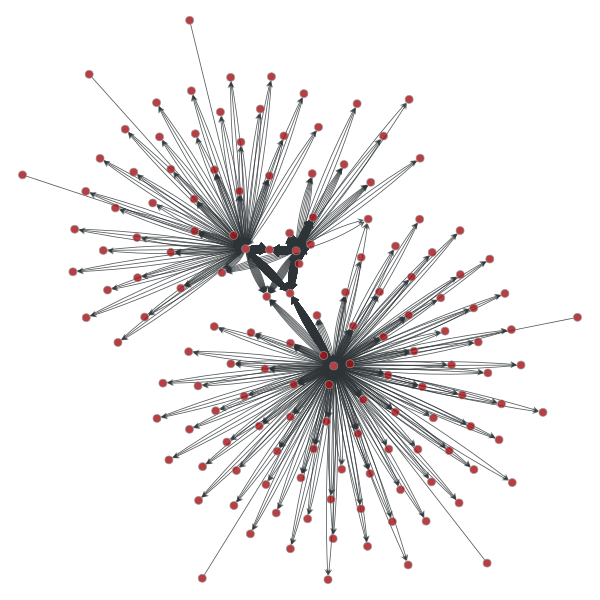}} &
\subfloat[t $\in$ [840 s, 960 s)]{\includegraphics[width = 1.4in]{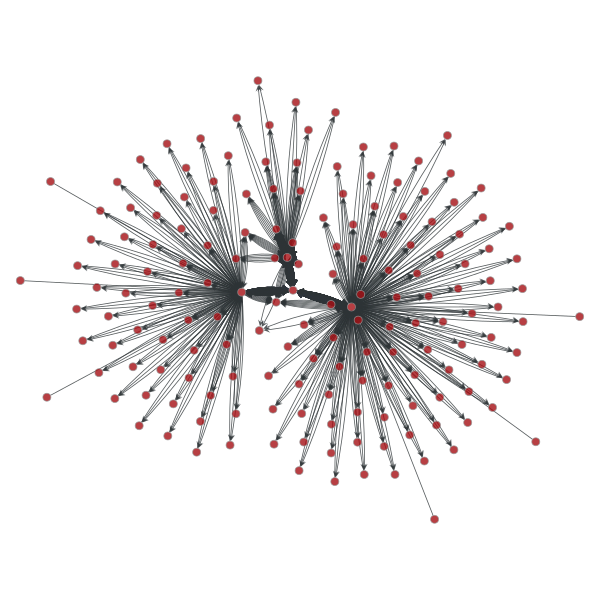}} &
\subfloat[t $\in$ [900 s, 1020 s)]{\includegraphics[width = 1.4in]{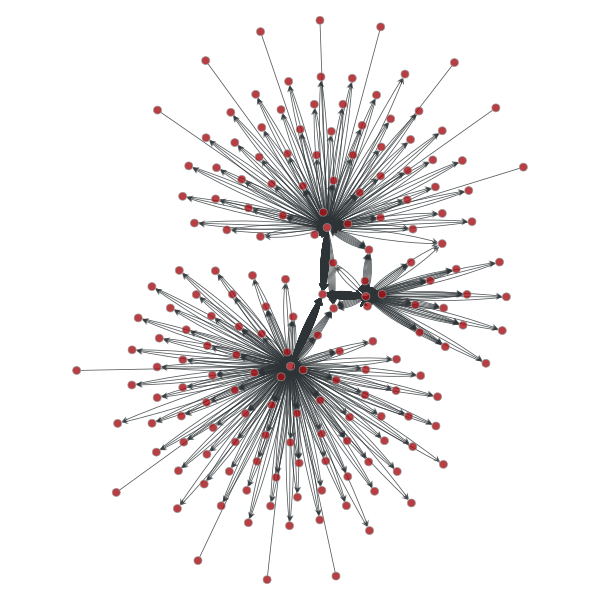}} \\
\subfloat[t $\in$ [960 s, 1080 s)]{\includegraphics[width = 1.4in]{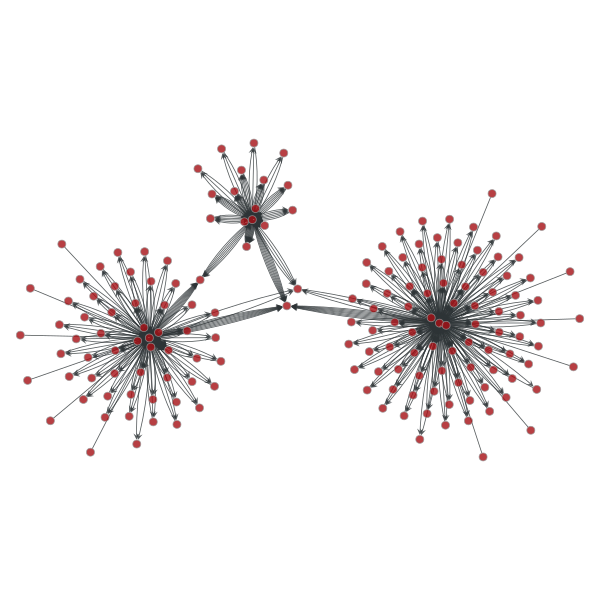}} &
\subfloat[t $\in$ [1020 s, 1140 s)]{\includegraphics[width = 1.4in]{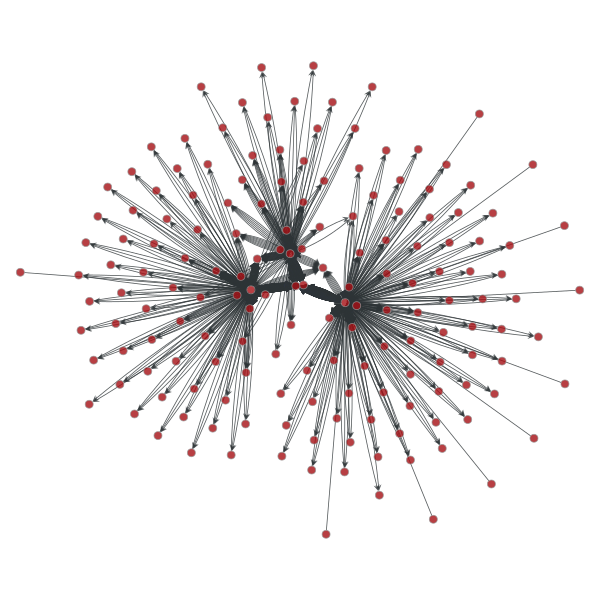}} &
\subfloat[t $\in$ [1080 s, 1200 s)]{\includegraphics[width = 1.4in]{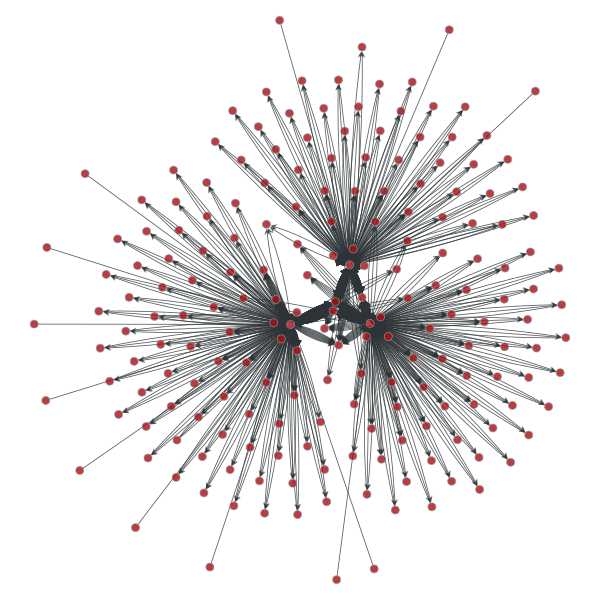}} \\
\subfloat[t $\in$ [1140 s, 1260 s)]{\includegraphics[width = 1.4in]{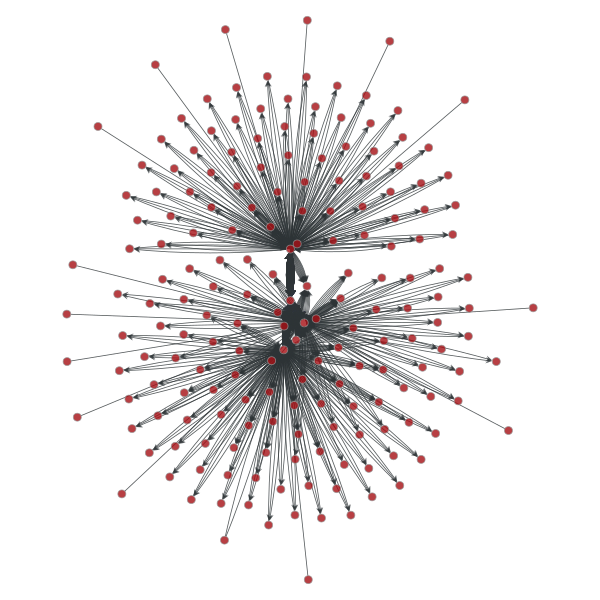}} &
\subfloat[t $\in$ [1200 s, 1320 s)]{\includegraphics[width = 1.4in]{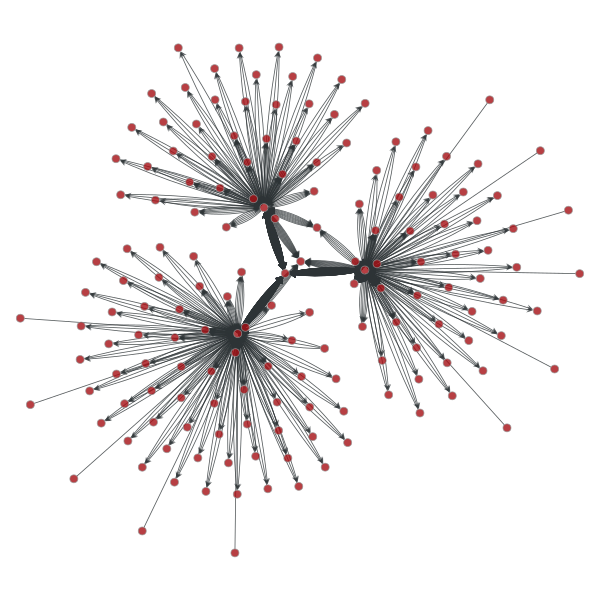}} &
\subfloat[t $\in$ [1260 s, 1380 s)]{\includegraphics[width = 1.4in]{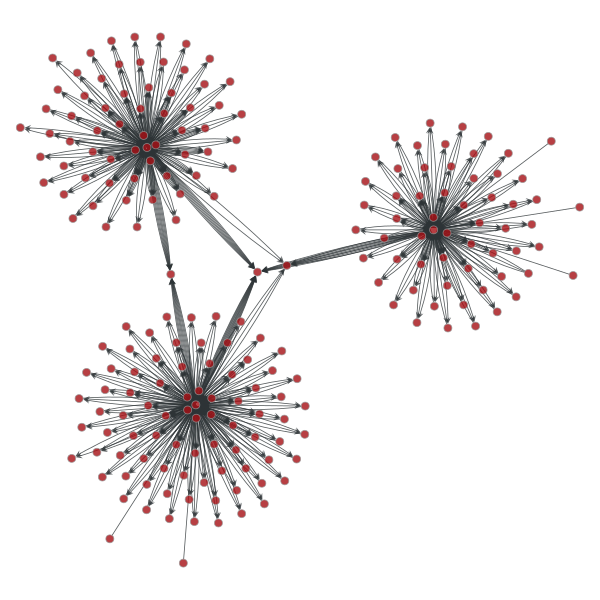}}
\end{tabular}
\caption{The series of graphs is generated from the packet capture file of scenario 12 (P2P botnet) in the CTU-13 dataset, trimmed to represent only packet transfers from one of its three malicious nodes. Interval size is set to 120 s and step size is set to 60 s.}
\label{fig:graph-sequence}
\end{figure}

Figure~\ref{fig:graph-sequence} shows the network communication graph over time for a set of malicious nodes from a P2P botnet. 
Each graph represents communication over a 120 second window. Note that we used a longer window to train our model, but the shorter window provides a clearer visual of the botnet evolution.
The P2P botnet shows a clear communication pattern over the time windows.
Our approach models the communication as a graph using graph-based features, and captures the evolution of the communication using an LSTM. Botnets characteristically have periodic behaviors and cycles of dormancy and activity~\cite{periodic}. 
Our intuition is that the graph-based features combined with an 
LSTM result in powerful classifier to detect botnet behavior.

\subsection{Algorithm Workflow}

\begin{figure}[!htbp]
\centering
\includegraphics[width=\textwidth]{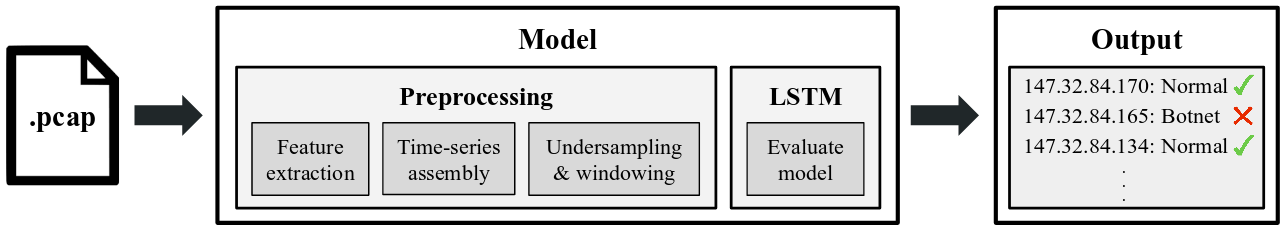}
\caption{Schematic of the Algorithm Workflow}
\label{fig:workflow}
\end{figure}

Figure~\ref{fig:workflow} shows the overall algorithm workflow. 
Our algorithm accepts network packet capture files (pcap) as input.
The packet captures contain a mix of botnet-infected and 
benign hosts (IP addresses), 
representing a good mix of botnet and normal network activity.
Each input 
packet capture file is first divided into overlapping time intervals,
each interval is then processed to extract graph features,
followed by feature normalization, and 
then assembled into a time-series as input to the LSTM.
Each time interval is labeled as malicious 
or benign, using ground-truth information available with the CTU-13 dataset.
During the model training phase, we additionally perform the steps 
of undersampling and windowing. 

During detection (or the testing phase), 
our input consists of network activity in the form of packet captures from each host.
The input is divided into time intervals similar to training,
followed by extraction of graph features, which are input 
to the trained LSTM model for classification
as 'botnet' or 'normal'.
We next describe the details of each step in the workflow.


\subsection{Data Preprocessing} 
\label{sec:data-processing}

\subsubsection{Graph-Based Feature Extraction and Normalization}

\begin{figure}[h]
\centering
\includegraphics[width=5in]{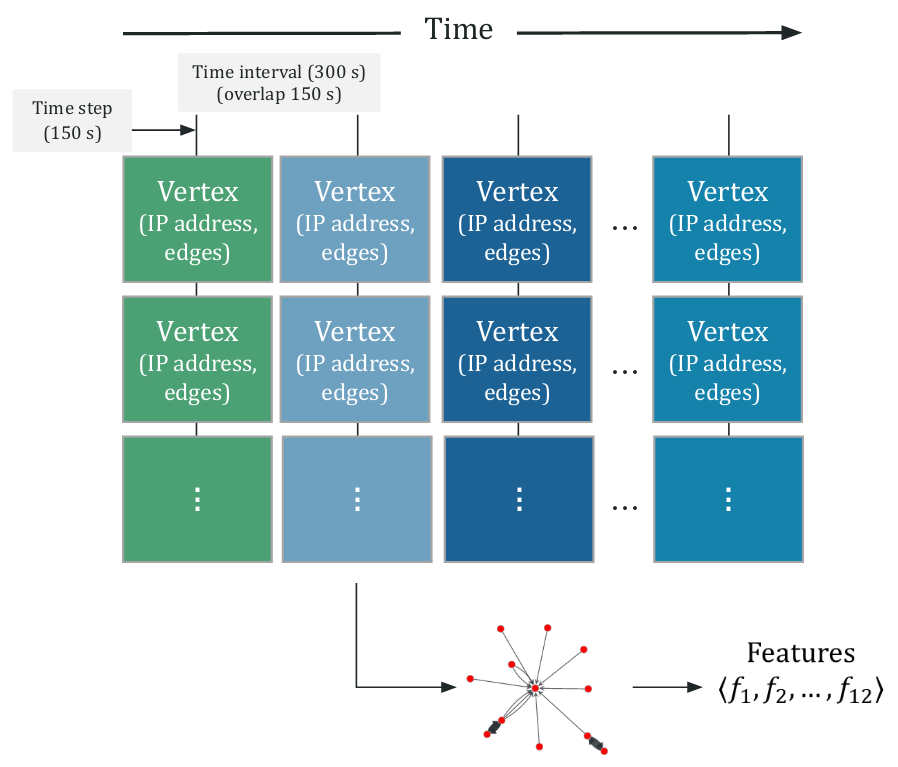}
\caption{Graph feature extraction from network data}
\label{fig:graphfeatures}
\end{figure}

The labeled input pcap files are divided into fixed-time, overlapping intervals; each interval is 300 seconds long and overlaps with 150 seconds of the previous interval. 
For each interval, a graph of the network is assembled, with IP addresses as nodes and individual packets as directed edges. 
For each node we extract ten graph features: out-degree, in-degree, out-neighbors, in-neighbors, PageRank centrality, betweeness centrality, eigenvector centrality, authority and hub centralities, and local clustering coefficient using the graph-tool library~\cite{graph-tool}. 
Then, per interval, features are normalized from $0.05$ and $0.95$ using the following equation:
$$ f_n = 0.05 + 0.9 * \frac{f - f_{min}}{f_{max} - f_{min}} $$
where $f_{min}$ and $f_{max}$ are the minimum and maximum values of the feature $f$ and $f_n$ is the normalized value of the feature $f$. 
The above normalization, similar to the one used by Guntuku et al.~\cite{Guntuku2013} was used because the sigmoid function is used to map to the output layer.
Figure~\ref{fig:graphfeatures} visually depicts the process of feature extraction across time windows. 

\subsubsection{Assembly of Time-Series}
Each IP address need not send or receive packets in every intervals, so we only extract features for nodes in some of the intervals. 
As we need a time-series for every node, we add a zero vector for every interval that a node is not present. 
Thus, the time-series for each node contains all the intervals in our graph, with zero padding for the intervals that the node does not receive or send packets.

\subsubsection{Undersampling and Windowing Intervals}
One of the main problems with cyber security datasets, including CTU-13, is that they are very unbalanced with far more negative (non-malicious) samples than positive (malicious) samples. 
Thus, we could not train a model directly on these time-series, since there were just a few (often under 5) malicious nodes with hundreds of non-malicious nodes. During the training phase,
we first undersampled the data, randomly discarding the data for many non-malicious samples, such that we retained ten non-malicious nodes for one malicious node (a 10:1 ratio). 
Then we divide each node's time-series of $x$ intervals into smaller windows of 5 intervals that overlap by 2 intervals (i.e. the first window contains intervals 1 - 5, the second window contains intervals 3 - 7, etc.). 
Thus, we end with a larger set of shorter time-series, still with a 10:1 ratio between non-malicious and malicious nodes.

\subsection{Long Short-term Memory (LSTM) Model}

We use a Long Short-term Memory (LSTM) as our time-series classification model. 
Botnets characteristically have periodic behaviors and cycles of dormancy and activity \cite{periodic}. 
Due to their layout, LSTM cells can remember values over arbitrary lengths of time. 
This lends LSTMs the power to recognize long-term periodic patterns, and 
makes them highly amenable for botnet analysis. 
We refer the reader to Hochreiter et al.~\cite{lstm} for a more detailed discussion of LSTM's.

\subsubsection{LSTM Structure}
We created the LSTM using the deep learning Keras framework \cite{keras}. 
The input layer consists of 10 neurons (one for each graph feature), the hidden layer consists of 64 neurons, and the output layer consists of a single neuron with sigmoid activation. 
We divided the data into training and testing sets by ensuring that $70\%$ of the positive (malicious) time-series are in the training set and the remaining $30\%$ in the testing set. 
The network was trained using the RMSProp optimizer~\cite{rmsprop} for 200 epochs. 
We used a weighted mean squared error loss function, weighting the loss of the malicious samples by six times (equivalent to oversampling our malicious time-series by six times) to account for our unbalanced data.

\subsubsection{Hyperparameters}
As data is in the form of a time-series, we capture the granularity of our time-series data in our hyperparameters. 
The only two hyperparameters in our model are \emph{window size} and 
\emph{step size}, which affect our time-series assembly phase. 
Note that exploring the effects of the values of these hyperparameters is prohibitively time-intensive, precluding any grid-search.
This would be the subject of our future research, but for this work
we picked reasonable values using visual exploration of the dataset,
and based on the lengths and densities of the packet capture files in the CTU-13 dataset, which range from approximately 15 minutes (scenario 11) to 5 hours (scenario 9).

\paragraph{Window size}
Window size is the duration of a segment of the packet capture that we use to construct a single graph; we use this graph to extract features for each node. It is important that this window size be large enough to capture trends in a single graph but also be small enough that the trends do not become too noisy. We set our window size to 300 s.

\paragraph{Step size}
Step size is the duration of increments between windows i.e. overlap = (window size) - (step size). Step size allows us to gradually transition our graphs so that we do not overlook  potential trends between but not within windows. For example, if a node suddenly increases and maintains high out-degree starting at time $t = 300$, the sudden change will not be realized if we have non-overlapping windows of size 300 s. We set our step size to 150 s.

\section{Evaluation and Results} 
\label{sec:results}

In this section, we first briefly discuss our evaluation dataset, 
then discuss our evaluation methodology and metrics, followed a discussion of the results.

\subsection{CTU-13 Dataset Description}
We trained and tested our models on the CTU-13 dataset. 
More specifically, we analyzed scenarios 6, 7, 10, 11 and 12 of the dataset which cover a variety of botnet characteristics (IRC, port scan, DDoS, and P2P). 
These scenarios span in total duration from 15 minutes to 5 hours (see Table \ref{table:ctu_13}), and the time that bots are active are slightly less than these times. 
Moreover, the number of infected nodes in these scenarios span from 1 to 10 machines. 
Note that scenarios 7 and 11 are especially short at 15 -- 25 minutes.
Every scenario includes a list of malicious hosts and the time at which they became infected. The data itself is available both as a packet capture file and a NetFlow file.

{
\renewcommand\arraystretch{1.25}
\begin{table}[!htbp]
\centering
\begin{tabular}{@{\extracolsep{4pt}}ccccccl}
\toprule
\multicolumn{1}{c}{\textbf{Scenario}} & \multicolumn{1}{c}{\textbf{Duration (hrs)}} & \multicolumn{1}{c}{\textbf{\# Packets}} & \multicolumn{1}{c}{\textbf{\# NetFlows}} & \multicolumn{1}{c}{\textbf{Size (GB)}} & \multicolumn{1}{c}{\textbf{\# Bots}} &
\multicolumn{1}{c}{\textbf{Characteristics}} \\
\midrule
6 & 2.18 & 38,764,357 & 558,920 & 30 & 1 & Port Scan \\
7 & 0.38 & 7,467,139 & 114,078 & 5.8 & 1 & HTTP \\
10 & 4.75 & 90,389,782 & 1,309,792 & 73 & 10 & IRC, DDoS \\
11 & 0.26 & 6,337,202 & 107,252 & 5.2 & 3 & IRC, DDoS \\
12 & 1.21 & 13,212,268 & 325,472 & 8.3 & 3 & P2P \\
\bottomrule
\end{tabular}
\caption{Statistics of the CTU-13 dataset scenarios that we analyzed \cite{ctu13}}
\label{table:ctu_13}
\end{table}
}

\subsection{Evaluation Methodology}

Due to the class imbalance between malicious and non-malicious nodes, train/test splits are applied independently to both the set of positive (malicious) and negative (non-malicious) examples e.g. given 100 positive examples (windows, as described earlier) and 1000 negative examples, a 70\%/30\% train/test split would reserve 70 positive examples and 700 negative examples for training and the remaining examples for testing. 
We evaluate our models in the following three ways.
\begin{enumerate}
    \item Train on 70\% of a scenario and test on the remaining 30\% of the same scenario
    \item Train on 100\% of a scenario and test on a different scenario
    \item Train on 70\% of a combination of scenarios and test on the remaining 30\% of the same combination of scenarios
\end{enumerate}

Results from evaluation 1 and 2 above are discussed in Section~\ref{sec:results_single},
while results from evaluation 3 are discussed in Section~\ref{sec:results_multiple}.

\paragraph{Metrics}
Generally in classification problems, the key base performance statistics are true positives (TP), true negatives (TN), false positives (FP), and false negatives (FN). 
Since the counts in themselves are not meaningful, detection literature sometimes analyzes their rates: true positive rate (TPR), true negative rate (TNR), false positive rate (FPR), and false negative rate (FNR).

More specifically in botnet literature, classification performances are evaluated via some or all of the following~\cite{flow-based}.
\begin{align*}
    Precision &= \frac{TP}{TP + FP} \\
    Recall &= \frac{TP}{TP + FN} \\
    F \textrm{-} measure &= 2 * \frac{recall * precision}{recall + precision}
\end{align*}

For botnet detection, we want to penalize false negatives more heavily than false positives but also need a manageable number of false positives that we can more closely analyze. 
Hence we seek a balance between TPR (sensitivity) and TNR (specificity), or equivalently TPR and FPR (the complement of TNR).
The ROC curve compares these two quantities at various thresholds, and thus gives a highly informative indication of the strength of a model. While precision-recall curves can be more informative in imbalanced class problems, 
there is a one-to-one correspondence between them~\cite{precision-recall-roc}. 
ROC curves traditionally have been used more often in these binary classification tasks, so we opt to examine them. 
Hence, we evaluate our performance by the area under the ROC curve (AUROC), where $0.5$ is random classification and $1$ is perfect classification.


\subsection{Results}


\subsubsection{Models trained on single scenario}
\label{sec:results_single}
We trained individual models on each of the scenarios we considered (6, 7, 10, 11 and 12),
and then evaluated it against different scenarios.
Further, we maintained a $70\%/30\%$ train/test split when evaluating on the same scenario as the train scenario. 
Table~\ref{table:single_model_roc_curve} summarizes the AUROC (area under ROC curve) values from the evaluation.

Note that most of the models at least perform well on the scenario they were trained on (see the diagonal values); scenario 7 is a clear exception, likely due to the very limited data.
Scenarios 7 and 11 are very limited datasets, as they contain less than 20 minutes of actual botnet activity, relative to the other datasets which are one or more hours long (see table \ref{table:ctu_13}). 
Moreover, the models generalize quite well to the other scenarios, especially the ones with less limited data (scenario 6, 10, and 12).


{
\renewcommand\arraystretch{1.25}
\begin{table}[!htbp]
\centering
\begin{tabular}{@{\extracolsep{4pt}}ccccccc}
\toprule
& \multicolumn{6}{c}{\textbf{Train Scenario}} \\ \cmidrule[.6pt]{2-7}
\textbf{Test Scenario} & 6 & 7 & 10 & 11 & 12 & Avg \\ \midrule
6 & 0.97 & 1.00 & 0.90 & 0.56 & 0.94 & 0.87 \\
7 & 1.00 & 0.67 & 0.87 & 0.58 & 1.00 & 0.82 \\
10 & 0.92 & 0.80 & 0.97 & 1.00 & 0.99 & 0.94 \\
11 & 0.99 & 0.89 & 0.96 & 0.97 & 0.96 & 0.95 \\
12 & 1.00 & 0.83 & 0.79 & 0.91 & 1.00 & 0.91 \\
\bottomrule
\end{tabular}
\caption{Table of AUROC (area under ROC) values of models trained on \emph{train scenario} and evaluated on \emph{test scenario}}
\label{table:single_model_roc_curve}
\end{table}
}

\subsubsection{Model trained on multiple scenarios}
\label{sec:results_multiple}

\begin{figure}[h]
\centering
\includegraphics[width=4in, trim={0cm 0cm 0cm 1.5cm},clip]{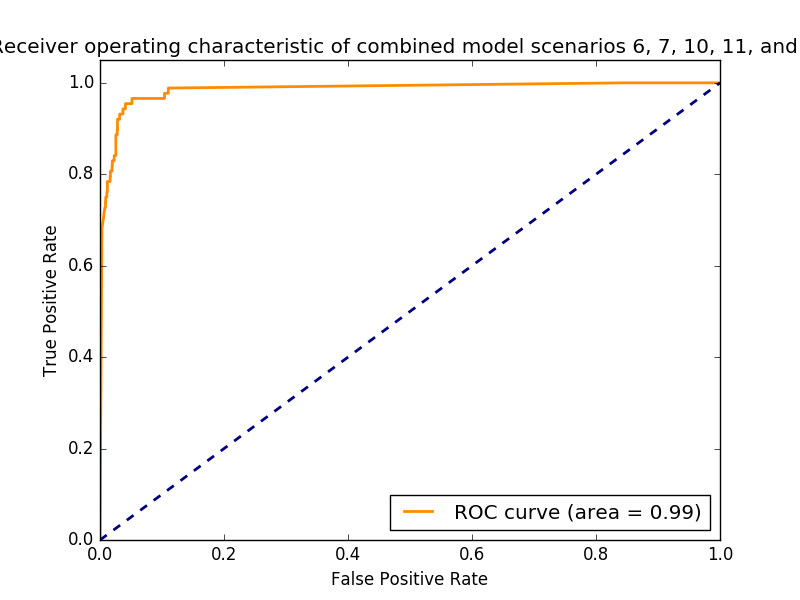}
\caption{ROC curve of model trained and tested on scenarios 6, 7, 10, 11, and 12}
\label{fig:combined_model_roc_curve}
\end{figure}

We next evaluated a model on a combination of scenarios 6, 7, 10, 11, and 12, using a
$70\%/30\%$ train/test split.
The model performed with an accuracy of $96.2 \%$, sensitivity (TPR) of $94.6 \%$, and specificity (TNR) of $96.3 \%$. 
Additionally, the false positive rate (FPR) was $3.73 \%$, and the false negative rate (FNR) was $5.43 \%$.
Figure~\ref{fig:combined_model_roc_curve} shows the corresponding AUROC.
Combining results from Section~\ref{sec:results_single}, and Section~\ref{sec:results_multiple}, we can conclude that our model 
generalizes strongly.  

\section{Discussion}
\label{sec:discussion}

In this section, we first discuss a comprehensive comparison of our approach and results with
other literature on the CTU-13 dataset~\cite{ctu13}.
We then briefly discuss our implementation
performance, followed by a discussion of limitations of our approach.

\subsection{Comparison of Results to Literature}
\label{sec:comparison}

Our approach achieves an
accuracy of $96.2 \%$, sensitivity (TPR) of $94.6 \%$, and specificity (TNR) of $96.3 \%$,
when evaluated on a combination of scenarios 6, 7, 10, 11, and 12.
We use these results for comparison with other approaches.

We note that the works cited below have evaluated their approaches on different subsets of scenarios from the CTU-13 dataset with different evaluation methodologies, thus performing a clear comparison with our approach is impossible. 
We nevertheless use the reported results as an indicator of performance, 
and carefully point out the differences with respect to our work.
Table~\ref{table:literature-analysis} summarizes the performance of different
approaches from the literature on the CTU-13 dataset.

García et al.~\cite{ctu13} tested a variety of models on their CTU-13 dataset. 
Their models' maximum accuracy rates were under 80\% for all the scenarios, including scenario 6, and generally the accuracies were far lower. 

Chowdhury et al.~\cite{graph-based} used unsupervised SOM clustering with graph-based features to filter the set of nodes into a significantly smaller cluster ($<1\%$ of the original set of nodes) of potentially malicious ones. 
While an effective filter tool, this approach is not a classification model or amenable to real-time analysis since it requires the entire dataset to cluster.
We thus don't report the metrics for this approach, but rather include it as an example of an unsupervised approach on the CTU-13 dataset.
We also note that the unsupervised nature of the algorithm means that it generalizes well. 

{
\renewcommand\arraystretch{1.25}
\begin{table}[h]
\centering
\captionsetup{justification=centering}
\begin{tabular}{@{\extracolsep{2pt}}lllcccc}
\toprule
\multicolumn{2}{c}{\textbf{Literature}} & \multicolumn{1}{c}{\textbf{Detection Method}} & \multicolumn{1}{c}{\textbf{Features}} & \multicolumn{1}{c}{\textbf{Detection Rate}} & \multicolumn{1}{c}{\textbf{Generalizes?}} &  \multicolumn{1}{c}{\textbf{\begin{tabular}[c]{@{}c@{}}Content- \\ agnostic?*\end{tabular}}} \\ \midrule
García et al. & \cite{ctu13} & \begin{tabular}[c]{@{}l@{}}anomaly detection,\\ behavioral\\ classification\end{tabular} & flow-based & $<0.80$ accuracy & No & No \\ \addlinespace[0.6em]
\begin{tabular}[c]{@{}l@{}}Chowdhury\\ et al.\end{tabular} & \cite{graph-based} & SOM clustering & graph-based & N/A & Yes** & Yes \\ \addlinespace[0.6em]
Torres et al. & \cite{rnn} & LSTM classification & \begin{tabular}[c]{@{}l@{}}flow-based /\\ behavioral\end{tabular} & \begin{tabular}[c]{@{}l@{}}$0.970$ TPR\\ within scenario,\\ $0.809$ TPR\\ across scenarios\end{tabular} & No & Yes \\ \addlinespace[0.6em]
Grill et al. & \cite{lams} & anomaly detection & flow-based & $0.94$ AUROC & No & No \\ \addlinespace[0.6em]
Terzi et al. & \cite{big-data} & \begin{tabular}[c]{@{}l@{}}anomaly detection\\ via k-means\\ clustering\end{tabular} & flow-based & \begin{tabular}[c]{@{}l@{}}$0.72$ TPR,\\ $0.987$ TNR,\\ $0.966$ accuracy\end{tabular} & Yes** & No \\ \addlinespace[0.6em]
Grill et al. & \cite{anomaly-combination} & \begin{tabular}[c]{@{}l@{}}anomaly detection\\ combinations\end{tabular} & various & \begin{tabular}[c]{@{}l@{}}$0.923$ precision,\\ $0.696$ accuracy\end{tabular} & No & No \\ \addlinespace[0.6em]
Yu at al. & \cite{session-based} & \begin{tabular}[c]{@{}l@{}}SDA-based\\ deep learning\\ classification\end{tabular} & flow-based & \begin{tabular}[c]{@{}l@{}}$>0.99$ accuracy,\\ precision, recall\end{tabular} & Not tested & No \\ \addlinespace[0.6em]
\begin{tabular}[c]{@{}l@{}}Haddadi\\ et al.\end{tabular} & \cite{choose-detection-system} & \begin{tabular}[c]{@{}l@{}}C4.5 decision tree\\ classification\end{tabular} & \begin{tabular}[c]{@{}l@{}}flow-based,\\ payload-based\end{tabular} & $>0.99$ TPR, TNR & Not tested & No \\ \addlinespace[0.6em]
Ryu et al. & \cite{ensembles} & \begin{tabular}[c]{@{}l@{}}GNB, NN, DT,\\ ensembles\end{tabular} & flow-based & $>0.99$ F1 score & Not tested & No \\ \addlinespace[0.6em]
Chen et al. & \cite{conversation-based} & \begin{tabular}[c]{@{}l@{}}Tree, forest\\ classifiers\end{tabular} & flow-based & \begin{tabular}[c]{@{}l@{}}$0.97$ accuracy,\\ $0.94$ TPR\end{tabular} & Not tested & No \\ \addlinespace[0.6em]
Sinha et al. &  & LSTM classification & graph-based & \begin{tabular}[c]{@{}l@{}}$0.962$ accuracy,\\ $0.956$ TPR,\\ $0.963$ TNR\end{tabular} & Yes & Yes \\
\bottomrule
\end{tabular}
\caption{Performance of approaches in literature on CTU-13 dataset}
\label{table:literature-analysis}
\vspace*{-10pt}
\begin{flushleft}
* We denote a detection method content-agnostic if it is port-, protocol-, and payload-agnostic \\
** These approaches use unsupervised learning and so naturally generalize across datasets
\end{flushleft}
\end{table}
}

Torres et al.~\cite{rnn} used LSTMs on simple, behavioral features, on scenario 1 and 6 of the CTU-13 dataset. For models trained and tested on the same scenario, their models achieved a TPR of 0.970 and TNR of 0.982. When applying a model trained on scenario 1, to scenario 6, they achieved a TPR of 0.809 and TNR of 0.970. While they achieve a high specificity in a given scenario, their sensitivity is significantly lower because the model generalizes poorly across scenarios.

Grill et al.~\cite{lams} developed an anomaly detection engine with LAMS (local adaptive multivariate smoothing), and tested on the CTU-13 datasets. 
They achieved a 0.94 AUROC on scenario 10 and a 0.87 and 0.88 AUROC on the other scenarios. 
While not a high accuracy score, incremental updates of the LAMS model allow for a better real-time model. 

Terzi et al.~\cite{big-data} used k-means clustering based on flow-based characteristics for anomaly detection. On scenario 10 of the CTU-13 dataset, they achieved  an accuracy of 0.966, TPR of 0.72, and TNR of 0.987. Despite a low TPR, they achieved a high accuracy for an unsupervised learning technique. 

Grill et al.~\cite{anomaly-combination} tested various convex combinations, which they prefer to arbitrary linear combinations for their increased interpretability, of anomaly detectors on the CTU-13 dataset. None of the combinations had both high precision and recall and in fact had a very low value for at least one or the other. Models with precision above 0.90 had recall under 0.10 and models with recall above 0.70 had precision under 0.30. The paper included a missing anomaly types (MAT) case, where some samples of malicious activity are absent from the training sample, which is a good measure of model generalization. In this case, the highest precision model had precision 0.923 and recall 0.057 and the highest recall model had precision 0.254 and recall 0.696. 

Chen et al.~\cite{conversation-based} extracted conversation, flow-based features, and trained several tree and forest classifiers. The models were trained on a combination of several scenarios from the CTU-13 dataset, and performed quite well. The random forest classifier performed the best, with an accuracy of approximately 0.97 and TPR of approximately .94.

Yu et al.~\cite{session-based} use unsupervised learning for feature extraction and then supervised learning via stacked denoising autoencoders (SDA) for classification. The paper creates a dataset by combining CTU-13 botnet and UNB ISCX IDS 2012 normal and malicious network traffic. 
The SDA approach works very well for both binary classification and multi-class classification, achieving over 0.99 accuracy, precision, and recall. The main drawback to the approach is that it requires examining and extracting features from the raw packet payloads.

Haddadi et al.~\cite{choose-detection-system} used the C4.5 algorithm to generate decision trees based on both a packet payload based system and a flow-based system. The packet-based system examines characteristics of the payload itself while the flow-based system examines more traditional flow statistics. The paper tests the models on all the CTU-13 scenarios and achieves very high TPR and TNR, often above 0.99 each. 

Ryu et al.~\cite{ensembles} trained Gaussian Naive Bayes, neural networks, decision trees, and various ensembles on flow-based features. Tested on scenario 4, 10, and 11 of the CTU-13 dataset, most of the methods perform very well, achieving an F1 score over 0.99.

To summarize, several of the approaches listed above seem to perform better than our approach in terms of the reported metrics, but those approaches also require access to content-specific features.
Our approach is by far the best in comparison with content-agnostic approaches, including several of the content-aware approaches.
As we discussed earlier, techniques such as encryption of the payload 
would defeat approaches depending on access to raw payload, while changes in botnet protocol such as payload length or packet inter-arrival times, could potentially defeat techniques which rely solely on flow-based features.


\subsection{Performance Analysis}
\label{sec:performance}

In this section, we briefly discuss the performance of our implementation.

\paragraph{Data processing}
Data processing involves (1) converting the packet capture file to its graph form, 
(2) extracting features, 
(3) assembling the time-series, and 
(4) post-processing (see section \ref{sec:data-processing}). 
Steps 1 and 4 are relatively cheap computationally. 
Step 3, while seemingly computationally simple, may be memory-intensive. This is because we must keep track of all nodes that were present in the packet capture file since we zero-pad them in intervals where they are absent.
Extracting features from the graphs (step 2), however, is by far the most time-consuming.

\paragraph{Theoretical time complexity analysis of graph feature calculations}
While not completely reflective of the time it takes to compute the features, a brief complexity analysis of our feature calculations gives meaningful information on the speed of feature extraction. 

Let $V$ and $E$ refer to the number of vertices and edges of a graph respectively.
Degree and neighbor feature calculations can be trivially done in $O(E)$. PageRank centrality uses an iterative update algorithm and waits for convergence; note that we set our convergence condition $\epsilon = 10^{-6}$. Betweeness centrality is $O(VE)$. Eigenvector centrality, Authority centrality, and Hub centrality use the power method and wait for convergence; note that we set our convergence condition $\epsilon = 10^{-6}$. Local clustering coefficient is $O(|V| \langle k \rangle^2)$, where $k$ is average out-degree. 
We recall that each of the above ten graph features need to be computed for each time interval during our training phase.
Please refer to the graph-tool library~\cite{graph-tool} for further details.

\paragraph{Empirical speed analysis of feature calculations}
We discuss our performance on scenario 10, which was by far the longest scenario we examined in the CTU-13 dataset. 
Empirically, the feature calculations using the power method predictably were significantly slower than the other methods due to their slow convergence. Due to the scenario's duration and our interval and step size, we construct 115 graphs for the packet capture file and calculate the graph features for each node. Extracting the graph-based features takes us 7000 minutes of computation time on a single-core machine. See section \ref{sec:limitations} for a discussion of speed limitations.

\paragraph{Training time}
Training the models based on the time-series input is far less time-intensive than the data processing phase. Training 200 epochs takes between several hours to less than a day. Given the size of the original packet capture file, these training times are reasonably fast.

\subsection{Limitations and Future Work} 
\label{sec:limitations}
The primary limitation with our approach was the sheer amount of time it takes to extract features. For scenario 10, the computation time to extract features given our computational resources exceeded the duration of the packet capture file.

This inefficiency derives mainly from our representation of the network as a multigraph i.e. we consider each packet transfer as an edge, resulting in many duplicate edges between pairs of hosts. 
We can condense these duplicates into a single weighted edge, like a flow. This simplification would yield many of the same feature calculations, and thus with no change to model accuracy, but at much higher throughput.
Our future work involves testing out the performance of our approach with this change.

Moreover, most of the graph-based feature algorithms are easily parallelizable, so making use of more cores would result in speed improvements. 
In addition, we ran our experiment on a CPU only; GPUs could definitely be used for speed improvements. Finally, memory access optimization's, especially in assembling the time-series, can also lead to speed improvements.

Another limitation of our current work, which we wish to improve in the near future, 
is the lack of evaluation of our model pre-trained on the CTU-13 dataset on other botnet or real-world datasets. 
We expect our generalizable approach to perform well across different datasets.
Further, we also wish to systematically diagnose our model's performance to gain a better understanding of its strengths and weaknesses.
For example, we wish to study why our model was better at detecting a particular botnet type over another type.

\section{Conclusions}
\label{sec:conclusion}

In this paper, we proposed a novel algorithm for botnet detection 
by tracking the evolution of a host's communication activity over time.
We combined graph-based features with a Long Short-term Memory (LSTM) neural network
classifier to detect malicious botnet hosts.
Our detection method is agnostic to both packet content and protocol 
and is thus hard to evade by techniques such as encryption.
We evaluated our approach on the CTU-13 dataset, 
and demonstrated an overall
accuracy of $96.2 \%$, sensitivity (TPR) of $94.6 \%$, and specificity (TNR) of $96.3 \%$,
when evaluated on a combination of scenarios 6, 7, 10, 11, and 12.
Moreover, our evaluation demonstrated that our approach generalizes 
across a variety of botnets.
Our biggest limitation currently is the time taken for pre-processing of data for training, but we believe this can be improved
dramatically by simplifying our graph construction methodology, and parallelizing feature extractions.

We also performed an exhaustive survey of the literature, and compared results from 
our approach with existing detection methods on the CTU-13 dataset.
We found that our approach is by far the best in comparison with content-agnostic approaches, including several of the content-aware approaches.
Overall, our approach complements the existing content-aware approaches.
We surmise that an ensemble of detectors consisting of the best performing content-aware approach, and our content-agnostic detector would result in a hard-to-evade botnet detection approach.


\section*{Acknowledgments}
A portion of the research described in this paper was carried out at the Jet Propulsion Laboratory, California Institute of Technology, under a contract with the National Aeronautics and Space Administration. 
The authors would
like to thank Dr. George Djorgovski and Stephanie Ding,
for discussions and feedback
that helped improve the ideas and methods expressed in this paper.

\bibliographystyle{unsrt}  
\bibliography{references}  

\begin{thebibliography}{10}

\bibitem{cryptocurrency}
Smominru.
\newblock
  \url{https://www.cyber.nj.gov/threat-profiles/botnet-variants/smominru}, Feb
  2018.

\bibitem{background}
R~Jaiswal and Shivraj Bajgude.
\newblock Botnet technology.
\newblock In {\em 3rd International Conference on Emerging Trends in Computer
  and Image Processing (ICETCIP’2013)}, pages 169--175, 2013.

\bibitem{signature-based}
Roberto Perdisci, Wenke Lee, and Nick Feamster.
\newblock Behavioral clustering of http-based malware and signature generation
  using malicious network traces.
\newblock In {\em Proceedings of the 7th USENIX Conference on Networked Systems
  Design and Implementation}, NSDI'10, pages 26--26, Berkeley, CA, USA, 2010.
  USENIX Association.

\bibitem{heuristic-based}
Kazuya Kuwabara, Hiroaki Kikuchi, Masato Terada, and Masashi Fujiwara.
\newblock Heuristics for detecting botnet coordinated attacks.
\newblock In {\em Proceedings of the 4th International Workshop on Advances on
  Information Security (WAIS2010)}, pages 603--607, 02 2010.

\bibitem{content-based}
W.~Timothy Strayer, David Lapsely, Robert Walsh, and Carl Livadas.
\newblock {\em Botnet Detection Based on Network Behavior}, pages 1--24.
\newblock Springer US, Boston, MA, 2008.

\bibitem{graph-based}
Sudipta Chowdhury, Mojtaba Khanzadeh, Ravi Akula, Fangyan Zhang, Song Zhang,
  Hugh Medal, Mohammad Marufuzzaman, and Linkan Bian.
\newblock Botnet detection using graph-based feature clustering.
\newblock {\em Journal of Big Data}, 4(1):14, May 2017.

\bibitem{ctu13}
Sebastian Garcia, Martin Grill, Jan Stiborek, and Alejandro Zunino.
\newblock An empirical comparison of botnet detection methods.
\newblock {\em computers \& security}, 45:100--123, 2014.

\bibitem{flow-based}
Matija Stevanovic and Jens Myrup~Pedersen.
\newblock An efficient flow-based botnet detection using supervised machine
  learning.
\newblock In {\em 2014 International Conference on Computing, Networking and
  Communications, ICNC 2014}, pages 797--801, 02 2014.

\bibitem{botgrep}
Shishir Nagaraja, Prateek Mittal, Chi-Yao Hong, Matthew Caesar, and Nikita
  Borisov.
\newblock Botgrep: Finding p2p bots with structured graph analysis.
\newblock In {\em Proceedings of the 19th USENIX Conference on Security},
  USENIX Security'10, pages 7--7, Berkeley, CA, USA, 2010. USENIX Association.

\bibitem{social-graph}
Jing Wang and Ioannis Paschalidis.
\newblock Botnet detection using social graph analysis.
\newblock {\em 2014 52nd Annual Allerton Conference on Communication, Control,
  and Computing, Allerton 2014}, 03 2015.

\bibitem{rnn}
Pablo Torres, Carlos Catania, Sebastián García, and Carlos Garcia~Garino.
\newblock An analysis of recurrent neural networks for botnet behavior
  detection.
\newblock In {\em Proc. IEEE Biennial Congr. Argentina (ARGENCON)}, 06 2016.

\bibitem{periodic}
Basil AsSadhan, Jose Moura, and David E.~Lapsley.
\newblock Periodic behavior in botnet command and control channels traffic.
\newblock In {\em GLOBECOM: Global Telecommunications Conference IEEE}, pages
  1--6, 11 2009.

\bibitem{graph-tool}
Tiago~P. Peixoto.
\newblock The graph-tool python library.
\newblock {\em figshare}, 2014.

\bibitem{Guntuku2013}
Sharath~Chandra Guntuku, Pratik Narang, and Chittaranjan Hota.
\newblock {Real-time Peer-to-Peer Botnet Detection Framework based on Bayesian
  Regularized Neural Network}.
\newblock jul 2013.

\bibitem{lstm}
Sepp Hochreiter and Jürgen Schmidhuber.
\newblock Long short-term memory.
\newblock {\em Neural Computation}, 9(8):1735--1780, 1997.

\bibitem{keras}
Fran\c{c}ois Chollet et~al.
\newblock Keras.
\newblock \url{https://keras.io}, 2015.

\bibitem{rmsprop}
T.~Tieleman and G.~Hinton.
\newblock {Lecture 6.5---RmsProp: Divide the gradient by a running average of
  its recent magnitude}.
\newblock Coursera: Neural Networks for Machine Learning, 2012.

\bibitem{precision-recall-roc}
Jesse Davis and Mark Goadrich.
\newblock The relationship between precision-recall and roc curves.
\newblock In {\em Proceedings of the 23rd International Conference on Machine
  Learning}, ICML '06, pages 233--240, New York, NY, USA, 2006. ACM.

\bibitem{lams}
Martin Grill, Tomás Pevný, and Martin Rehak.
\newblock Reducing false positives of network anomaly detection by local
  adaptive multivariate smoothing.
\newblock {\em Journal of Computer and System Sciences}, 83, 04 2016.

\bibitem{big-data}
Duygu Sinanc, Ramazan Terzi, and Seref Sagiroglu.
\newblock Big data analytics for network anomaly detection from netflow data.
\newblock In {\em Computer Science and Engineering (UBMK), 2017 International
  Conference on. IEEE}, pages 592--597, 10 2017.

\bibitem{anomaly-combination}
Martin Grill and Tomás Pevný.
\newblock Learning combination of anomaly detectors for security domain.
\newblock {\em Computer Networks}, 107, 06 2016.

\bibitem{session-based}
Yang Yu, Jun Long, and Zhiping Cai.
\newblock Session-based network intrusion detection using a deep learning
  architecture.
\newblock In {\em Modeling Decisions for Artificial Intelligence, vol. 10571 of
  Lecture Notes in Computer Science}, pages 144--155, 09 2017.

\bibitem{choose-detection-system}
Fariba Haddadi, Duong-Tien Phan, and A.~Nur Zincir-Heywood.
\newblock How to choose from different botnet detection systems?
\newblock {\em NOMS 2016 - 2016 IEEE/IFIP Network Operations and Management
  Symposium}, pages 1079--1084, 2016.

\bibitem{ensembles}
Songhui Ryu and Baijian Yang.
\newblock A comparative study of machine learning algorithms and their
  ensembles for botnet detection.
\newblock {\em Journal of Computer and Communications}, 06:119--129, 01 2018.

\bibitem{conversation-based}
Ruidong Chen, Weina Niu, Xiaosong Zhang, Zhongliu Zhuo, and Fengmao Lv.
\newblock An effective conversation-based botnet detection method.
\newblock {\em Mathematical Problems in Engineering}, 2017:1--9, 04 2017.

\end{thebibliography}

\end{document}